\documentclass{aa}  
\usepackage{graphicx}
\usepackage{epsfig}
\usepackage{txfonts}
\usepackage[dvips]{color}

\begin{document}

\title{Reaffirming the connection between the Galactic stellar
warp and the Canis Major overdensity}
   \subtitle{}
   \author{M. L\'opez-Corredoira\inst{1}, Y. Momany\inst{2}, 
   S. Zaggia\inst{2}, A. Cabrera-Lavers\inst{1,3}}
   \offprints{martinlc@iac.es}

\institute{
$^{1}$ Instituto de Astrof\'\i sica de Canarias, C/.V\'\i a L\'actea, s/n,
E-38200 La Laguna (S/C de Tenerife), Spain \\
$^{2}$ INAF-Oss. Astronomico di Padova, Vicolo dell'Osservatorio 5,
35122 Padova, Italy \\
$^3$ GTC Project Office, C/.V\'\i a L\'actea, s/n,
E-38200 La Laguna (S/C de Tenerife), Spain \\
}

   \date{Received May 10 2007; accepted July 24 2007}

  \abstract
   {}
%
%
%
{We aim to understand the real nature
of  the  stellar overdensity  at southern   galactic  latitudes in the
region of CMa.}
%
%
{We perform a critical re-analysis and discussion of recent results
presented in the literature which interpret the CMa overdensity as the
signature of an accreting dwarf galaxy or a new substructure
within the Galaxy. Several issues are addressed.}
%
%
%
{We show that arguments against the ``warp'' interpretation are based
on an erroneous perception of the Milky Way.
There is nothing anomalous with colour--magnitude
diagrams on opposite sides of the 
average warp mid-plane being different. 
We witnessed the rise and fall of
the blue plume population, first attributed to young stars in a disrupting
dwarf galaxy and now discarded as a normal disc population.
Similarly, there is nothing anomalous in the outer thin+thick disc
metallicities being low ($-1<$ [Fe/H] $<-0.5$), and spiral arms (as part
of the thin disc) should, and do, warp. Most importantly, we
show unambiguously that, contrary to previous claims, the warp 
produces a stellar overdensity that is distance-compatible with 
that observed in CMa.}
%
%
%
{The CMa over-density remains fully accounted for in a  first
order approach by Galactic models without new substructures. 
Given the intrinsic uncertainties
(concerning the properties of the warp, flare and disc cutoff, the
role  of extinction and  degeneracy), minor deviations with respect to
these  models  are not  enough  to  support  the hypothesis of an
accreted dwarf  galaxy or  new  substructure within the  Milky Way
disc.}
   
\titlerunning{The Warp-CMa connection}
\authorrunning{L\'opez-Corredoira et al.}
   \maketitle
%

\section{Introduction}

Since its discovery (Martin   et al.  \cite{mart04}), the Canis  Major
(CMa)  over-density of stars has been  the subject of  a lively debate
over whether it is a dwarf galaxy or simply the warped/flared Galactic
disc  [see    Momany    et  al.       \cite{moma06} (hereafter   M06),
L\'opez-Corredoira
\cite{lope06} (hereafter L06), and  references therein].  However, there are
many alternative ``solution''  papers (for and against the dwarf
galaxy  origin). In the absence of a clear-cut  evidence in  favour of  
an extra-Galactic origin (e.g.    chemical enrichment), attention was  focused on the star
counts and   stellar  populations of the  CMa   over-density.  In this
regard, the  new wide-field  surveys by  Conn et al.  (\cite{conn07},
hereafter C07), Butler et al.   (\cite{butl07}, hereafter B07) and  de
Jong    et   al.  (\cite{jong07},     hereafter  d07)   presented deep
colour--magnitude diagrams (CMDs) that, in principle, challenge the warp
hypothesis.  In this paper, we explain why this is not the case.

We do not explain all the second order details of the CMa overdensity
solely in terms of the warping/flaring of the Milky Way stellar disc.
Momany et al. (\cite{moma04}), M06 and L06 demonstrated that, on the
basis of its star counts, the CMa over-density cannot unambiguously be
disentangled from the warp feature, and that a Galactic origin, given
the uncertainties in all Galactic models, remains the first-order
explanation.  Our purpose is to reply to the claims of C07, B07 and
d07, critically re-analysing some of their results and conclusions, in
the light of the ``warp'' solution and demonstrating that it is still
the most plausible one to explain the CMa stellar overdensity.

\section{Critical re-analysis of the CMa overdensity}

In the following subsections we show that the observational data of C07, B07 and d07
can  be better explained  by  normal Milky  Way warped
disc modelling than  by a  dwarf galaxy  or  new  substructure in  the
Galaxy. In particular, erroneously interpreted or wrongly used aspects
will be pointed out.

\subsection{CMDs on opposite sides of the warp}

The main result of the C07 and B07 surveys (and major objection to the
warp interpretation)  is that CMDs  of the centre of CMa and control northern
hemisphere fields show  different morphology and  star counts.  Their
large survey  coverage allowed a  CMD comparison not only for opposite
hemisphere fields at different latitudes  but also for fields  equally
distant  from the nominal  warped mid-plane.  The  details of this CMD
comparison   are found   in   B07   (\S  7.2,   Fig.   5)
[$b\approx+8^\circ  $ $vs$ $b\approx-15^\circ   $] and C07 (\S  6.2.1,
Fig.   28) [$b\approx+4^\circ$ $vs$  $b\approx-9^\circ$].  Both groups
attribute    the differences  to  the presence   of  an extra  stellar
population (in addition  to that of the   warp).  B07  and C07's
conclusions were based on the finding by M06 (\S 4.2) that the mid-plane
for  Galactic longitudes  of $l\sim  240^\circ  $ should be on average
$b\sim  -3^\circ $. However, the  presence of a  plane  of symmetry at
$b\sim -3^\circ$ is not what M06 meant. Moreover, the well-known
age--metallicity--distance  degeneracy   can  also introduce some  minor
quantitative error   in any   interpretation of   CMa CMDs. In  the
following paragraphs  we   show  why  a  CMD comparison at   two given
latitudes is not straightforward.

\begin{figure}
{\par\centering \resizebox*{8cm}{8cm}{\includegraphics{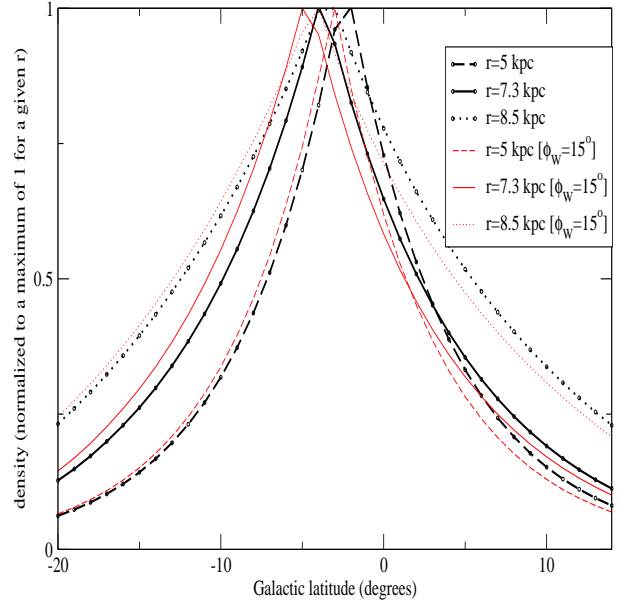}}\par}
\caption{Stellar density (normalized to a maximum of unity)
as a function of galactic latitude for $l=240^\circ$ and source
distance at 5.0, 7.3 and 8.5 kpc respectively, according to the L06
Galactic model (with $\phi _W=+5^\circ $), or the L06 model with $\phi
_W=+15^\circ $ (as derived by M06). This figure shows that
intercepting the disc at symmetric latitudes with respect to the local
mid-warp, one still obtains stellar densities with different
profiles.}
\label{Fig:lop06}
\vspace{-4mm}
\end{figure}

It is crucial  to understand that the  asymmetries  introduced by the
warp change for different values of  distance from the Sun ($r$),
and that this  implies different densities  as a  function  of the absolute
magnitudes of the  main  sequence stars.  To  quantify  this effect we
make use of the L06 warp model\footnote{The L06 warp model is the same
as the L\'opez-Corredoira et al. (\cite{lope02}) model, except for a $\phi
_W$ of $+5^\circ $  and an extrapolated  $z_W(13\ {\rm kpc}<R<16\ {\rm
kpc})=z_W(13\ {\rm  kpc})$ for the  southern warp.}  and  calculate in
Fig.~\ref{Fig:lop06} the stellar  density [normalized to unity in each
case] as a function of  Galactic latitude for different distances.
If we use a line of nodes angle of  $\phi _W= 15^\circ $ (M06) instead
of $\phi _W=5^\circ $  (L06) the results remain  similar except for  a
slight shift of 1--2 degrees in the latitude of maximum density.  Thus,
Fig.~\ref{Fig:lop06} shows that the warp asymmetries are a function of
the heliocentric distance and  thereby neither the overall star counts
nor the general features of the CMD are expected  to be similar in the
$b\approx+8^\circ $   and $b\approx-15^\circ $  or $b\approx+4^\circ $
and $b\approx-9^\circ$ diagrams.

Indeed, one should keep in mind  that any point (colour--magnitude pair)
of a CMD displays an  integration of all detected stellar  populations
along the line of  sight (i.e. at  all distances) for each  CMDs pair.
One    cannot  isolate   (via  visual inspection or isochrone
superposition)    the  $\sim7.2$  kpc  stellar  populations   in these
diagrams.  The isochrone plotting in Figure~28  of C07 may show a
higher  stellar density at   distances of $\sim7.2$ kpc. Nevertheless,
presumed CMa sequences may include stars at nearer distances and
one cannot  isolate the CMa populations from  young,  faint and nearby
main-sequence   stars.    This  precludes  any quantitative
conclusion  on the  non-similarity  of  CMDs  pairs at  CMa distances.
Integrating along the $b\approx-15^{\circ}$ line  of sight one clearly
``travels'' below the stellar disc. However, assuming  that the disc is
warped  downwards in these directions, one will sample more disc
stars  and might  even intercept the  warping disc  at a certain scale
height.  This is the opposite for the $b\approx+8^{\circ}$ field where,
integrating along this line of sight, one will sample less and less of
the warped disc   and, most importantly,   will {\em not} intercept  a
significant portion of its scale height in the III quadrant. The
sudden  appearance  of  a seemingly separated  main  sequence  in  the
$b\approx-15^{\circ}$ field is possibly the signature of the warp.

One should also keep in mind the unknown but important uncertainties
due to possible variations of the reddening law while crossing the
warped mid-plane at large distances.  The left panel of Fig.~9 in B07
shows two vertical lines that delimit the ``high reddening region'':
$E_{(B-V)}\ge 0.30$. This sketches the strong downward disc warping as
traced by the interstellar dust.  This dust asymmetry of $\sim
2$--$3^{\circ}$ is consistent with that found for the gas and stars
(M06).  Interestingly, a hint that the stellar disc shows a $\sim
2$--3$^{\circ}$ downward bending in this line of sight is found by
B07.  The total extinction for $b\approx+8^\circ $ and
$b\approx-15^\circ $ or $b\approx+4^\circ $ and $b\approx-9^\circ$ is
similar, however the differential reddening is expected to be quite
different (again, because the lines of sight follow very different
paths): there is a different reddening as a function of the magnitude
that therefore produces different apparent shapes of the main
sequence.

Thus, there is nothing anomalous  in CMDs being equally displaced in
latitude  from  the $b\approx    -3^\circ$  warped  mid-plane being
different: integration along the line of sight is not symmetric, and
the warp is not asymmetric for all heliocentric distances, and 
neither is extinction.

\subsection{Distance of the maximum overdensity produced by the warp}

B07 make use of a modified L06 warp model\footnote{They take constant
height at $R>14$ kpc instead of $R>13$ kpc, and $\phi _W=-5^\circ $
instead of $\phi _W=+5^\circ $.  These parameters, however, do not
affect the present comments.}  and infer that the peak of the star
counts, for a given population of stars along the CMa line of sight,
is at $(m-M)\approx 10.5$ (i.e.  a distance $d=1.3$ kpc).  This value
is at odds with a correct application of L06, and implies a serious
analysis error by B07.

Leaving aside the  extinction,  the star counts   for a
stellar population  with  magnitude $M$  up  to $m$, having  a density
distribution of  $\rho   (\vec{r})$ and  certain $(l,b)$ Galactic
coordinates  within an area of $\omega  $ radians can  be expressed as
follows:
\begin{equation}
N(m)=\omega \int _{0}^{r(m)}\rho [x,l,b]x^2dx
,\end{equation}\[
r(m)=10^{[m-M+5]/5} {\rm \ (distance)}
,\]
rather than $N(m)=\omega \int _0^r \rho xdx$, as used by B07
(Pe\~narrubia, priv. comm., Oct.-2006, who also performed the B07
calculations), based on equations (7) and (8) of Pe\~narrubia et
al. (\cite{pena05}).  We believe that the Pe\~narrubia formulation is
incorrect because the counts per magnitude interval are not equal to
the counts per unit distance [it is not correct to use $N(m)=N(r)$ in
equation (7) of Pe\~narrubia et al. (\cite{pena05})].  Thus, it is
straightforward to obtain:
\begin{equation}
\frac{dr(m)}{dm}=\frac{ln\ 10}{5}r(m)
,\end{equation}
and consequently the counts at magnitude $m$ per unit magnitude are
\begin{equation}
A(m)\equiv    \frac{dN(m)}{dm}=\frac{ln\ 10}{5}\omega \rho
[r(m),l,b]r(m)^3 
,\end{equation}  
rather than $\frac{dN}{dm}\propto\rho r$ as used in B07 (Pe\~narrubia,
priv. comm. Oct.-2006).  Should one apply this last (incorrect)
formula, one would derive a distance of the maximum warp overdensity
at $(m-M)\approx 11.1$.  There is an $(m-M)\approx 0.6$ difference
between our reconstruction of the erroneous calculations of B07 and
their reported value, and this is probably due to the inclusion of
extinction in the B07 calculations. B07's erroneous calculation most
probably stems from this wrong application of the stellar statistics
equation [see also errors listed in L06(\S 2.3)].

\begin{figure}
{\par\centering \resizebox*{8cm}{8cm}{\includegraphics{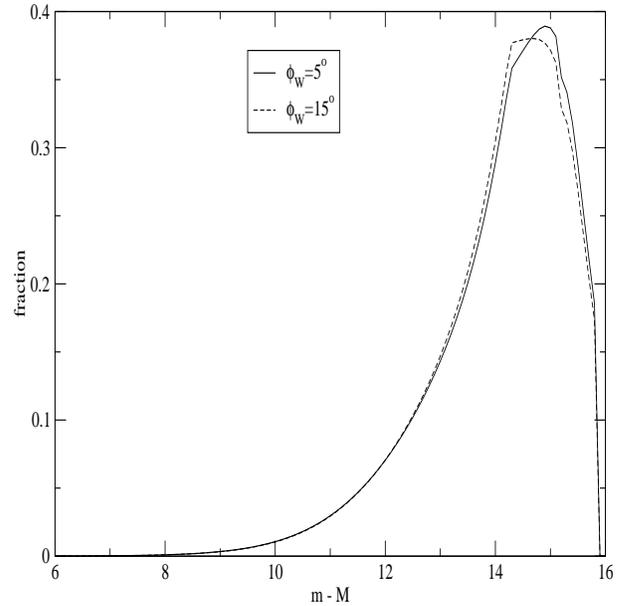}}\par}
\caption{The fraction of stars (normalized to give an area of unity) as a
function  of  $(m-M)$ for $(l,b)=(241.5^\circ,-7.5^\circ)$  predicted by 
 the  L06 warp-model  (assuming zero extinction  along the  line of
sight). The short-dashed line is obtained with a modified L06 model with
$\phi _W=15^\circ $.}
\label{Fig:but14}
\vspace{-4mm}
\end{figure}

Not surprisingly, several authors (Bellazzini et al. \cite{bell06b};
M06; L06) have already demonstrated that the correct distance is in
the range 5--10 kpc.  Figure~\ref{Fig:but14} shows our application of
the L06 warp model and shows a clear disagreement with Fig.~14 of B07.
In our case, and assuming zero extinction, the maximum of the star
counts [proportional to $\rho (r)r^3$] is found at
$(m-M)_{\circ}\approx 14.9$, a distance of $\sim 10$ kpc. Indeed, this
result is implicit in L06 (Fig. 2): since the peak of the red clump
stars was found around $m_K=13.3$, this corresponds to $(m-M)\approx
15$ (considering that $M_{K,red\ clump}\approx -1.65$).  On the other
hand, the total extinction for the CMa centre is around $A_B=0.99$
(Schlegel et al. \cite{schl98}). When this is taken into account, the
resulting maximum should be at $(m_B-M_B)>13.9$, depending slightly on
the absolute magnitude $M_B$ of the adopted population.

Our Fig.~\ref{Fig:but14} shows that the B07 star-count maximum (their
Fig.~14) is in agreement with the L06 warp model predictions. There is
a small difference concerning the depth of the main sequence
star-count maximum (FWHM).  This is predicted around $\sim2.2$ mag
(see also L06[\S 2.3; Fig.  3]) whereas the B07 observed depth is 30\%
lower (L06, \S 2.3).  This is a natural effect of inaccuracies in the
warp+flare modelling.  One should, however, keep in mind that
extinction would contribute towards narrowing the FWHM, thus reducing
the disagreement between the warp model expectations and
observations. This is particularly true if an important fraction of
the total extinction, along the line of sight, is associated with
distances near the CMa overdensity.  Either way, the warp
model-observation differences are too small to justify the exclusion
of the warp hypothesis and the need for an extra (accreted)
population.

One might wonder why we argue that our first-order Galactic model
solves for the distribution of the observed overdensity while we argue
that other models like Besan\c con (Robin et al.  2003, extensively
used by C07 and B07) do not.  The Besan\c con model has been tested
(C07 and B07) and provides an excellent description of the Galaxy in
many lines of sight. But the Besan\c con model does not include a
lower amplitude of the southern stellar warp with respect to the
northern one (L06).  Moreover, the model does not include the CMa
young population of stars associated with the Norma--Cygnus spiral
arms (see next subsection).  Therefore, distentangling different
stellar sequences in CMa CMDs (via galactic model comparisons) is
intrinsically difficult, even before investigating the presence of a
dwarf galaxy.

\subsection{CMa blue plume population}

Colour--magnitude diagrams of the centre of CMa show the conspicuous
presence of stars brighter than the old MS turnoff.  Bellazzini et al.
(\cite{bell04}) were the first to interpret this population as
$\sim$1--2 Gyr CMa stars, arguing that the CMa stellar overdensity
(originally identified by older red clump stars) also shows an
overdensity of young stars.  However, the same blue plume population
was explained (Carraro et al. \cite{carr05}; Moitinho et al.
\cite{moit06}) in terms of a much younger ($\sim100$ Myr) thin disc
spiral arm and inter-spiral population.  The lack of a blue plume
population was noted in all the B07 fields {\em above} the mid-plane,
and in fields between $193^{\circ}\le~l~\le 220^{\circ}$ and {\em
below} the mid-plane.  This leaves little space for a unique CMa blue
plume population connection and, admittedly, C07 favour a Galactic
origin of the blue plume population.

An erroneous approach of B07 or d07 is to make a distinction between
the warp+flare (M06, L06) and the spiral arm (Carraro et
al. \cite{carr05} and Moitinho et al. \cite{moit06}) scenarios. The
warp+flare and spiral arm interpretations are complementary, as they
refer to (young and old) stellar populations having the same {\em
Galactic} origin.  The Milky Way disc populations, regardless of their
age, {\em do} warp.  Spiral arms warp downwards in the 3rd quadrant of
the external disc.  Consequently, an excess of blue plume stars
appears when one observes along southern Galactic latitude lines of
sight.  Indeed, the blue plume population survey by Moitinho et
al. (\cite{moit06}, the lower panel of their Fig. 2) shows a clear
sign of warping for the young stellar populations in relation to the
CMa distance.  A successive wide-field III quadrant survey by Carraro
et al. (\cite{carr07}) confirms the presence of a warped young stellar
populations at about $\sim 9$ kpc, associated with the Norma--Cygnus
spiral arm. Even more interestingly, Carraro et al.  (\cite{carr07})
trace the presence of an older ($\sim 7$ Gyr) population and identify
its MS and red giant stars.  They conclude that this $\sim7$ Gyr
population is associated with the old and warped thin/thick disc
component.

Thus, the ``CMa blue plume population'' is most likely to reflect a
{\em warped} Milky Way thin disc population, as those with older ages;
no conspiracy of two different effects as said by B07 or d07, just
warping disc populations.

\subsection{CMa extension, disruption and tidal effects}

In their figure~9, B07 shows that there is little surface density
gradient in longitude across their CMa survey area.  In particular,
there is a clear density profile compatibility between the red clump
(based on 2MASS star counts by Bellazzini et al. \cite{bell06b}) and
the old main sequence (based on the B07 optical survey) studies.
A near-flat profile  over  $\sim30^{\circ}$ implies a higher  elongation
and weakens the definition of an  ``overdensity''.  This is {\em
not} a  minor  detail in   the CMa  debate   because the  CMa  stellar
populations  (thought of as an  extra population)   would actually  {\em
connect up} with the Argo overdensity (Rocha-Pinto et al.
\cite{roch06}: $l\sim300^{\circ}$).    
If one accepts an  extra-Galactic origin  for such  a  huge and  elongated
overdensity (covering an extension larger  than  the entire III
quadrant) it is legitimate to ask: where is the ``Galactic'' stellar
disc?

For the old CMa main sequence population, B07 estimate a
FWHM$_{(l,b)}=(>27^{\circ},\sim6^{\circ})$ and a ratio of $>5:1$.  On
the other hand, for the young CMa main sequence population they derive
FWHM$_{(l,b)}=(>20^{\circ},\sim2^{\circ})$ and a ratio of $>10:1$,
i.e. the most recent stars formed at the centre would have been the
last to be stretched, which is against what is expected in a dwarf
galaxy.  Interestingly, the young main sequence density profile (B07,
Fig.~9, right panel) shows a peak in their longitude coverage.  This
can easily fit the scenario in which the Galactic stellar disc (in the
CMa line of sight) follows the trend of the warped gas: starts warping
downwards, reaches a maximum at Galactocentric distances of $\sim 13$
kpc ($\sim$ the CMa distance), and then re-approaches the nominal
Galactic mid-plane. Should this be the case, then the peak in the
young main sequence density profile at $240^{\circ}$ would correspond
to the ``interception'' of the thin disc warping downwards.

The effect of a dwarf galaxy or massive substructure with an extension
of several kpc embedded in the Milky Way at only 13 kpc of the
Galactic centre should present distortions of the Galactic disc and
spiral arms, which are not observed.  Moreover, and as discussed in
B07, while orbiting the Galactic centre, a disrupting coplanar
satellite would be subject to continuous tidal forces that give rise
to tidal tails.  Any bound portion should not, however, show a flat
density distribution, as found for CMa.

\subsection{The CMa metallicity}

Recently, D07 presented a quantitative analysis of wide-field CMDs in
and around the CMa centre. Their isochrone fitting technique suggests
a metallicity, [Fe/H], of between $-0.6\pm 0.3$ for the fields that
are further away from the plane and $-1.0\pm 0.2$ (D07; table 2) for
the fields that are closer to the plane (and consequently with larger
problems with extinction). Bellazzini et al.  (\cite{bell06a}) with a
similar method had reported $-0.7<[M/H]<-0.4$ near the centre of the
Canis Major overdensity.  Even a higher metallicity,
$[Fe/H]\simeq0.3$, has been reported by Sbordone et al. (2005) for one
single CMa candidate.  D07 also anticipate FLAMES and AAOMEGA
spectroscopic metallicity results (by Martin et al., in prep.) of
kinematically selected members showing [Fe/H] $\approx -0.9$.  This
leads D07 to argue against the warp/flare hypothesis, with which we do
not agree.

\begin{figure}
{\par\centering \resizebox*{8cm}{8cm}{\includegraphics{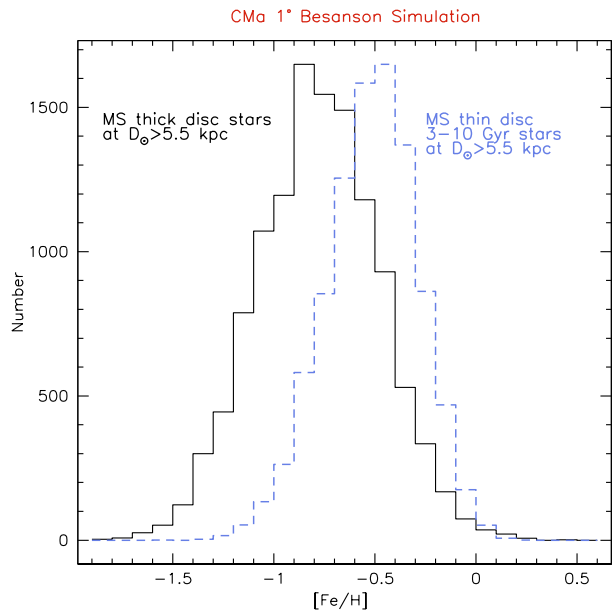}}\par}
\caption{A $1^{\circ}$ Besan\c con simulation along the CMa centre line of
sight    showing  the  expected     thin    and thick Galactic    disc
metallicities. The dashed  histogram (main  sequence thin disc  stars)
has been normalized to that of the thick disc.}
\label{Fig:besancon}
\vspace{-4mm}
\end{figure}

Leaving aside the uncertainties (rough extinction assumptions,
foreground/background contamination and distance--metallicity
degeneracy) that affect the D07 analysis, it is interesting to note,
again, that there is nothing strange in this ``low'' metallicity
within our Galaxy.  This is shown in the bimodal distribution of
Figure~\ref{Fig:besancon}, the expected Milky Way thin and thick disc
metallicities (from a Besan\c con simulation for old main sequence
stars; Robin et al. \cite{robi03}).  From a more general point of
view, Hammer et al.  (\cite{hammer07}) observe that the chemical
abundance of the Milky Way outskirts is three times lower than those
of most spiral galaxies within a similar mass range. Added to a
smaller stellar mass and angular momentum, Hammer et al.  infer an
exceptionally quiet formation history for the Milky Way, apparently
escaping any significant merger over the past $\sim 10$ Gyr.  Chemical
abundances (Sbordone et al. 2005) suggest that some CMa regions might
have experienced some unusual star formation, but it is at present
inconclusive about CMa origin.

\section{Discussion and conclusions}

We re-examined the most recent claims (C07, B07 and D07) in favor of
an accreted dwarf galaxy in Canis Major.  In particular, we
unambiguously show (Fig.~\ref{Fig:but14}) that the Galactic warp
produces an overdensity whose maximum coincides in distance with that
reported for CMa. We show that contrary claims were based on erroneous
calculations.  We also argue that the warp and the spiral
arms/segments scenarios should {\em not} be put forward as two
different explanations.  Young and old Galactic components, equally,
warp.  Lastly, we argue that a low metallicity for CMa populations is
not at odds with typical thick disc populations.

As it stands, {\em we cannot see any property of the CMa stellar
overdensity that cannot be accounted for in terms of a ``smoothly''
warped and flared Galactic stellar disc}.  On the contrary, the lack
of ancient stellar populations\footnote{The small number of detected
RR Lyrae stars in the CMa over-density can be accounted for by the
halo and thick disc RR Lyrae population (Mateu et
al. \cite{mate07}).}, a compatibility of CMa radial velocities, proper
motion, and chemical composition with typical disc values indicate the
Galactic origin of the overdensity.

 Acknowledgments:  Thanks are given to the referee Blair Conn
(ESO) for very helpful comments. M. L\'opez-Corredoira was supported
by the {\it Ram\'on y Cajal} Programme of the Spanish Science
Ministry. This research has partially been supported by the
Italian INAF PRIN grant CRA 1.06.08.02.

\end{document}